\documentclass[journal]{IEEEtran}%,draftcls,onecolumn
\usepackage{graphicx}

\begin{document}
%
% paper ttle
\title{Adiabatic quantum computation with flux qubits,\\
first experimental results.}

\author{S.H.W.~van~der~Ploeg, A.~Izmalkov, M. Grajcar, U. H\"ubner, S. Linzen,
S. Uchaikin, Th. Wagner, A.~Yu.~Smirnov, A. Maassen van den Brink,
M.H.S. Amin,A.M. Zagoskin, E.~Il'ichev and H.-G. Meyer% <-this % stops a space
\thanks{SvdP, AI and EI were supported by the RSFQubit and EuroSQIP
projects, MG by Grants VEGA 1/2011/05 and APVT-51-016604.}%
\thanks{S.H.W.~van~der~Ploeg, A.~Izmalkov, U. H\"ubner, S. Linzen, Th. Wagner,
E.~Il'ichev and H.-G. Meyer are at the Institute for Physical High Technology,
P.O. Box 100239, D-07702 Jena, Germany}% <-this % stops a space
\thanks{M. Grajcar is at the Department of Solid State Physics,
Comenius University, SK-84248 Bratislava, Slovakia}%
\thanks{S. Uchaikin, A.~Yu.~Smirnov, A. Maassen van den Brink and M.H.S. Amin
are at D-Wave Systems Inc., 100-4401 Still Creek Drive, Burnaby, B.C., V5C 6G9 Canada}%
\thanks{M.H.S. Amin and A.M. Zagoskin  are at the Physics and Astronomy Dept.,
The University of British Columbia, 6224 Agricultural Rd.,
Vancouver, B.C., V6T 1Z1 Canada}%
\thanks{A. Maassen van den Brink and A.M. Zagoskin  are presently also at:
Frontier Research System, RIKEN, Wako-shi, Saitama, 351-0198, Japan}}%

% use only for invited papers
\specialpapernotice{(Invited Paper)}

% make the title area
\maketitle

\begin{abstract}
Controllable adiabatic evolution of a multi-qubit system can be used
for adiabatic quantum computation (AQC). This evolution ends at a
configuration where the Hamiltonian of the system encodes the
solution of the problem to be solved. As a first steps towards
realization of AQC we have investigated two, three and four flux
qubit systems. These systems were characterized by making use of a
radio-frequency method. We designed two-qubit systems with coupling
energies up to several kelvins. For the three-flux-qubit systems we
determined the complete ground-state flux diagram in the three
dimensional flux space around the qubits common degeneracy point. We
show that the system`s Hamiltonian can be completely reconstructed
from our measurements. Our concept for the implementation of AQC, by
making use of flux qubits, is discussed.

\end{abstract}

%\begin{keywords}
%flux qubits, coupling \textsf{check wether we need this}
%\end{keywords}
% Note that keywords are not normally used for peerreview papers.

% For peer review papers, you can put extra information on the cover
% page as needed:
% \begin{center} \bfseries EDICS Category: 3-BBND \end{center}
%
% For peerreview papers, inserts a page break and creates the second title.
% Will be ignored for other modes.
\IEEEpeerreviewmaketitle

\section{Introduction}
% The very first letter is a 2 line initial drop letter followed
% by the rest of the first word in caps.
%
% form to use if the first word consists of a single letter:
% \PARstart{A}{demo} file is ....
%
% form to use if you need the single drop letter followed by
% normal text (unknown if ever used by IEEE):
% \PARstart{A}{}demo file is ....
%
% Some journals put the first two words in caps:
% \PARstart{T}{his demo} file is ....
%
% Here we have the typical use of a "T" for an initial drop letter
% and "HIS" in caps to complete the first word.
\PARstart{T}{he} concept of quantum computation by adiabatic
evolution, i.e. Adiabatic Quantum Computation (AQC), was introduced
by Farhi and coauthors \cite{Farhi2000}. It is an alternative to the
``standard'' concept of quantum computation where one has to
construct a universal set of gates with long coherence times. As the
latter requirement is not easy to achieve in superconducting quantum
bits (qubits), presently the longest times achieved are of the order
of several $\mathrm{\mu s}$ and only achieved at an optimal point
\cite{Bertet2005b, Yoshihara2006}, AQC seems to be better as the
system is only required to stay in the ground state. This led
Kaminsky and co-authors to propose a scalable superconducting
architecture for adiabatic quantum computation \cite{Kaminsky2004},
where they make use of 3 Josephson Junction persistent current
qubits \cite{Mooij1999}.

In the AQC concept the solution of some non-polynomially (NP) hard
problem is encoded into the ground state of a complex multi-qubit
Hamiltonian $H_\mathrm{P}$. The parameters of such a Hamiltonian can
be controlled continuously so in this sense this method is similar
to \textit{analog} computation. The computation itself contains
three majors steps: (A) preparation of the system in a well-known
state with Hamiltonian $H(t=0)=H_\mathrm{I}$ at time $t=0$, (B)
adiabatic evolution during the time interval $t_\mathrm{calc}$
towards the problem Hamiltonian $H_\mathrm{P}$ and (C) readout of
the ground state of the problem Hamiltonian, which gives the answer.
Therefore, in order to build a quantum computer based on AQC one has
to be able to construct a qubit system with a controllable
Hamiltonian and one has to be able to read out this system once the
problem Hamiltonian $H_\mathrm{P}$ is encoded into it.

\begin{figure}
  % Requires \usepackage{graphicx}
 \centering  \includegraphics[angle=270,width=8cm]{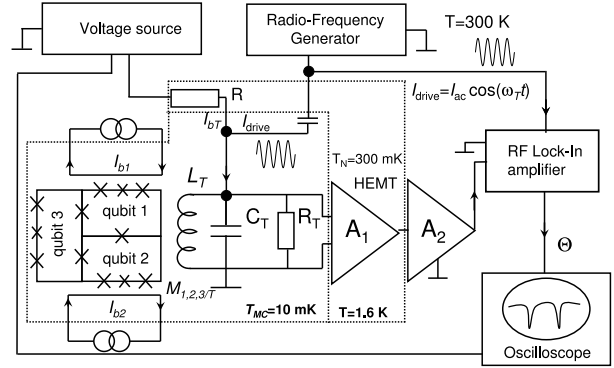}\\
\caption{Experimental setup for the demonstration of adiabatic
quantum computation in a three-qubit-system, see section
\ref{sec:AQC}. The qubits are coupled to a resonant tank circuit for
the readout. The flux bias to the qubits is provided by a dc current
$I_{bT}$ applied to the coil $L_T$ and bias currents $I_{b1,2}$
applied to additional bias wires, allowing full control over the
individual qubits fluxes $f_i$. In the case of more qubits, more
current sources would be used.}\label{setup}
\end{figure}

There has been an experimental implementation of an AQC algorithm in
an NMR three-qubit quantum computer by Steffen \textit{et al.}
\cite{Steffen2003}. By using their computer they solve the so-called
MAXCUT problem. This problem is one of the Non-Polynomially hard
problems of Graph theory where the object of the problem is to find
the maximal ``cut'' of a graph \cite{Garey1976} and is
mathematically equivalent to finding the ground state of some
Ising-type Hamiltonian. Some of the authors have proposed how to
implement and readout such an algorithm making use of flux qubits in
a setup like the one shown in Fig.\ref{setup} \cite{Grajcar2005}.

In this paper we review our recently obtained results from coupled
flux qubits using our and implemented method for their
characterization. Also we will show some of our results on multi
qubit systems.

\section{Coupling of qubits}

In order to implement any protocol for quantum information
processing, qubits should be coupled. Three Josephson junction
persistent current flux qubits \cite{Mooij1999} can be coupled
inductively just by placing them next to each other, however the
achievable coupling strength is rather weak \cite{Izmalkov2004a}. A
larger coupling strength can be achieved by making use of the
kinetic inductance of a common leg between them \cite{Majer2005}.
Even stronger coupling can be achieved by inserting a Josephson
junction in this common leg \cite{Levitov2001}. The latter also
provides the possibility of ferromagnetic coupling, whereas the
other methods only provide anti-ferromagnetic coupling
\cite{Grajcar2005a}. By insertion of an additional coupler loop
between the qubits, one could realize controllable/switchable
coupling between them \cite{Ploeg2006}. This is particulary
important for the realization of scalable superconducting
architecture for AQC. Controllable coupling enables one to construct
a Hamiltonian where both the energy bias $\epsilon_i$ and the
coupling terms $J_{ij}$ can be set \textit{in-situ}.

%\begin{figure}
%  % Requires \usepackage{graphicx}
%   \centering \includegraphics[width=7cm]{EnergyBW_FM}\\
%\caption{The first two energy levels for an FM-coupled two qubit system.
%Calculated for $I_{p,1}=150~\mathrm{\mu A}$, $I_{p,2}=100~\mathrm{\mu A}$,
%$\Delta_1=\Delta_2=100~\mathrm{mK}$ and J=-300~mK}\label{EnergyFM}
%\end{figure}
A coupled $N$-flux-qubit system can be described by the
Hamiltonian:
\begin{equation}\label{eq_H_q} H_q =-\sum_i^N [
\epsilon_i  \sigma_{z}^{(i)} + \Delta_i \sigma_{x}^{(i)}] +
\sum_{i<j}J_{ij}\sigma_{z}^{(i)}\sigma_{z}^{(j)}.
\end{equation}
Here $\Delta_i$ is the tunneling amplitude, or half energy
splitting, of qubit $i$ at $f_i=0$ and
$\varepsilon_{i}=\Phi_0I_{p,i}f_i$ gives the energy bias applied to
qubit $i$ in terms of its persistent current $I_{p,i}$ and the
normalized flux bias $f_i=\Phi_{x,i}/\Phi_0-1/2$ ($\Phi_0$ is the
flux quantum). The symbols
$\sigma_{x}^{(i)},\sigma_{y}^{(i)},\sigma_{z}^{(i)}$ denote the
Pauli matrices of the \textit{i}-th qubit. The coupling energy is
given by:
\begin{equation}\label{J}
    J_{ij}=\left(M_{ij}\pm\frac{\Phi_0}{2\pi
    I_{c,ij}}\right)I_{p,i}I_{p,j},
\end{equation}
where $M_{ij}$ represents both the magnetic inductance between
qubits $i$ and $j$, as well as the kinetic inductance of a common
leg. The second term describes the coupling due to an additional
junction with critical current $I_{c,ij}$ inserted in the common leg
between both qubits. The many-qubit system described by
Eq.~\ref{eq_H_q} can be characterized by the set of eigenstates
$|\mu\rangle$ and eigenenergies $E_{\mu}, \mu = 1,.., 2^N,$ which
can be obtained from the solution of the equation: $H_q |\mu \rangle
= E_{\mu} |\mu \rangle.$ As an example the first two levels $E_0$
and $E_1$ are shown in Fig.~\ref{EnergyAFM} for a two qubit system
with anti-ferromagnetic coupling ($J_{12}>0$).

%
%\begin{equation}\label{QubitHamiltonian}
 %   H_q=\sum_i^n
  %  \Delta_i\sigma_x^{(i)}+\Phi_0I_{p,i}f_i\sigma_z^{(i)}+
   % \sum_{j<i}J_{ij}\sigma_z^{(i)}\sigma_z^{(j)}
%\end{equation}
\section{Qubit readout using a low-frequency resonator}\label{Sec:theory}
For the readout of the qubit system we use a high-quality LC
resonator (tank-circuit) consisting of a superconducting coil with
inductance $L_T$ in parallel with a capacitance $C_T$
\cite{Ilichev2004}. In order to analyze the response of such a tank
to the qubits we have to consider the complete Hamiltonian of the
qubits system and readout resonator: $H = H_q + H_T
+H_\mathrm{int}$, where $H_T$ is the Hamiltonian of the driven tank
circuit and $H_\mathrm{int}$ the interaction term between tank and
qubits. In principle, one should also add a term containing the
dissipative interaction with the environment, which would result in
line broadening, $\Gamma_{\mu \nu}$, of the many-qubit spectrum.
However, the measurements are performed at a resonator frequency
$\omega_T$ which is much less than the energy spacings, $\omega_{\mu
\nu} = E_{\mu \nu}: \omega_T \ll |\omega_{\mu \nu}|$, of the coupled
qubits. Under such non-resonant conditions the effects of
dissipation on the measurements are proportional to the small ratio
$\Gamma_{\mu \nu}/|\omega_{\mu \nu}|$, and can be neglected.

The driven tank-circuit Hamiltonian can be written in terms of the
non-commuting current, $I_T$ and voltage, $V_T$ operators. (here
$[V_T, I_T]_-= -i \hbar \omega_T^{~2}, $ with $\omega_T =
1/\sqrt{L_T C_T}$ the resonance frequency) as:
\begin{equation} \label{eq_H_T}
H_T = \frac{C_T V_T^2}{2} + \frac{L_T I_T^2}{2} - L_T I_T
I_\mathrm{drive},
\end{equation}
where $I_\mathrm{drive}(t)=I_{ac}\cos(\omega t)$ is the driving
current of the tank. The interaction term is given by
$H_\mathrm{int}= - I_T \Phi$ with $\Phi$ the magnetic flux through
the tank loop created by all qubits, $\Phi = \sum_{i} \lambda_i
\sigma_{z}^{(i)}$. $\lambda_i = M_{iT}I_{p,i}$ is the coupling
parameter between the $i$-th qubit and the tank, which is
proportional to the  mutual inductance between qubit and the coil of
the tank-cirquit $M_{iT}$ and to the magnitude of the persistent
current of the qubit. By using the Heisenberg equations $i \hbar
\dot{V}_T = [V_T, H]_-$ and $i \hbar \dot{I}_T = [I_T, H]_-$ we
obtain the following equation for the averaged current and voltage
in the LC-circuit: $\langle \dot{I}_T \rangle = \langle V_T \rangle
/L_T, $
\begin{equation} \label{eq1_V_T}
\left( \frac{d^2}{dt^2} + \gamma_T \frac{d}{dt} + \omega_T^2 \right)
\langle V_T \rangle = \frac{1}{C_T} \dot{I}_\mathrm{drive} +
\omega_T^2 \frac{d}{dt} \langle \Phi_\mathrm{tot}(t) \rangle,
\end{equation}
where the damping of the tank $\gamma_T=\omega_T/Q_T^{~0}$ was added
on an \textit{ad-hoc} basis. The flux $\Phi_\mathrm{tot}$
incorporates not only the flux $\Phi$ which the qubits would apply
without their interaction with the tank, but also includes the
response due to the tank according to:\begin{equation}
\label{eq2_Phi_Tot}
 \Phi_{tot} = \Phi (t) + (i/\hbar) \int_{-\infty}^{t} dt_1 [\Phi(t),\Phi(t_1)]_ - I_T(t_1).
\end{equation}
Here we have used the following notations for the (anti)commutator
of two operators A and B: $[A,B]_{\pm} = AB \pm BA,$ and the
interaction picture.

Eq. \ref{eq1_V_T} has the solution $ \langle V_T(t) \rangle  = V_T
\cos(\omega t + \Theta)$ with $\Theta$ the phase shift of the
voltage relative to the driving current. For the case of resonant
driving, $\omega = \omega_T$, the voltage-current phase shift is
determined by the sum, over all qubit combinations, of the real part
of the qubit susceptibility $\chi_{ij}'$:
\begin{equation} \label{eq1_shift}
\tan \Theta = - \frac{Q_T}{L_T} \sum_{ij} \lambda_i \lambda_j
\chi_{ij}'(\omega_T).
\end{equation}

\begin{figure}\centering
  % Requires \usepackage{graphicx}
  \includegraphics[width=6cm]{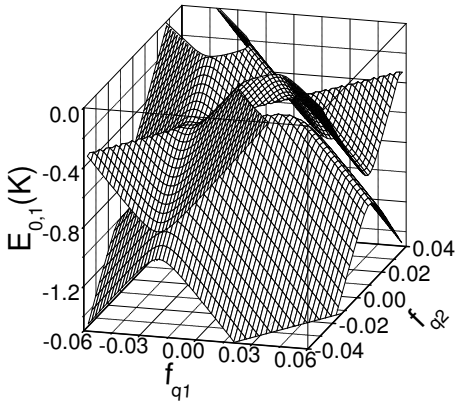}\\
\caption{The first two energy levels for an AFM-coupled two qubit
system. Calculated for $I_{p,1}=100~\mathrm{nA}$,
$I_{p,2}=150~\mathrm{nA}$, $\Delta_1=\Delta_2=100~\mathrm{mK}$ and
J=300~mK.}\label{EnergyAFM}
\end{figure}

\begin{figure}\centering
  % Requires \usepackage{graphicx}
  \includegraphics[width=6cm]{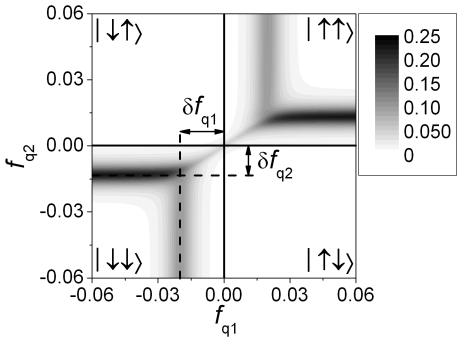}\\
\caption{$-\tan\Theta$ for an AFM-coupled two qubit system. The tank
response is calculated for the same qubit parameters as used in
Fig.~\ref{EnergyAFM}. The read-out parameters are $L_T=80$~nH,
$M_{1,T}=M_{2,T}=50$~pH and $Q_T=500$ with an effective temperature
$T=70$~mK. }\label{AFresponse}
\end{figure}

The many-qubit susceptibility at the frequency of the tank
$\omega_T, \omega_T \ll (E_{\mu}-E_{\nu})/\hbar$ is calculated in
the appendix \ref{suceptibility} and results in:
\begin{equation} \label{eq2_shift}
\chi_{ij}'  = \sum_{\mu \neq \nu} \frac{\rho_{\mu} - \rho_{\nu}}
{E_{\mu} - E_{\nu}} \langle \mu |\sigma_{z}^{(i)} |\nu \rangle
\langle \nu | \sigma_{z}^{(j)} |\mu \rangle .
\end{equation}
Here $\rho_\mu, \rho_\nu$ denote the equilibrium occupation of the
qubit energy eigenstates: $\rho_{\mu} = e^{-E_{\mu}/k_B
T}/(\sum_{\alpha}e^{-E_{\alpha}/k_B T})$. A numerical calculation of
the predicted response for a 2 qubit system is shown in
Fig.~\ref{AFresponse}. By comparing this with the energy diagram in
Fig.~\ref{EnergyAFM} it is seen that the predicted signal is largest
close to the anticrossings where the ground state $E_0$ has the
largest curvature due to the superposition of flux states.

\section{Experimental Implementation}

%In a radio-frequency method analyzed above the tank circuit plays
%the role of parametric transducer~\cite{Grajcar2005}. Indeed any
%variation of real part susceptibility $\chi_{ij}$ of a qubits
%system is transduced to changes of the tank resonant frequency.
For our experiments we fabricate the qubit-system of interest inside
a prefabricated superconducting planar coil made out of niobium on a
Si chip. This coil forms the tank-circuit together with an external
capacitor~\cite{May2003}. On the coil chip there are also Nb lines
which allow the application of an asymmetric flux bias to the
qubits. By applying a dc bias current $I_{bT}$ through the tank and
currents $I_{b1}$ and $I_{b2}$ (in the 3 qubit example shown in
Fig.~\ref{setup},~\ref{AQCreadout}a) through the lines we can
separately change the magnetic flux in the qubits. For more qubits
one should add additional lines.

As shown above, measuring the shift of the resonance frequency
allows one to read-out the total susceptibility of the qubit system.
Therefore, the tank-circuit is driven with a current
$I_\mathrm{drive}(t)=I_{ac}\cos(\omega t)$ at a frequency $\omega$
close to the resonance frequency $\omega_T$ of the tank circuit. The
variation of the resonance frequency, due to the change in the
susceptibility of the qubit system according to (\ref{eq1_shift}),
can be measured by detecting the phase shift $\theta$ between the
$\textit{ac}$-current $I_\mathrm{drive}$ applied through the coil
and $\textit{ac}$--voltage over the tank circuit. This is realized
by using the setup shown in Fig.~\ref{setup}. The voltage across the
tank circuit is amplified by a cold HEMT-based amplifier followed by
a room temperature amplifier. The phase shift $\Theta$ is measured
using a lock-in amplifier with part of the signal which is used to
bias the qubit used as a reference.

%In order to reconstruct $\Theta$ the tank circuit is driven by an
%$ac$ current $I_\mathrm{drive}$  A $dc$ bias current $I_{bT}$ is
%also applied to the tank circuit to change a dc flux of a qubit
%systems. Both currents produce the total applied magnetic flux to
%the qubits $\Phi_e = \Phi_\mathrm{dc} +\Phi_\mathrm{rf}\cos \omega
%t$.

In the simplest case - one qubit at low temperature $k_BT \ll
(E_{\mu}-E_{\nu})/\hbar$, $\Phi_{tot}$ is proportional (through the
mutual inductance) to the persistent current $I_q$ flowing in the
qubit system. Since a qubit is in the ground state $E_0$ the current
$I_q$ is proportional to $\frac{dE_{{0}}}{d\Phi_{x}}$. By taking
into account Eq.~\ref{eq1_V_T} it can be easily shown that a shift
of the tank resonant frequency due the tank-qubit interaction (and,
therefore, $\tan\Theta$ or $\chi_{ij}$) is proportional to the
ground state curvature $\frac{d^2E_{0}}{d\Phi_{x}^2}$. Because the
ground state curvature is maximal near the anticrossing, the output
signal as a function of the applied magnetic flux undergoes a narrow
dip~\cite{Greenberg2002} at $\Phi_0/2$. From the shape of the dip
the one-qubit Hamiltonian can be reconstructed by making use of the
following equations \cite{Grajcar2005}:
\begin{equation} \label{dip1}
I_p =
\frac{\tan\theta_{max}\Phi_{0}F_{HW}}{k^{2}QL_{q}2\sqrt{2^{2/3}-1}},
\end{equation}

\begin{equation} \label{dip2}
\Delta = I_p\frac{\Phi_{0}F_{HW}}{2\sqrt{2^{2/3}-1}}.
\end{equation}
At higher temperatures one can fit the dip directly to the theory as
given by Eqs.~(\ref{eq1_shift}) and (\ref{eq2_shift}).

By properly biasing a multi qubit system one can use this single
qubit reconstruction procedure for a multi qubit system. The bias to
such an $N$ qubit system  is applied as follows: for $N-1$ qubits
$\epsilon_i \gg \Delta_i$ and $\epsilon_i \gg J_{i,j}$ for any $j$.
Then the $\Delta_i$ in Eq.~(\ref{eq_H_q})  for $N-1$ qubits and also
the interaction between all qubits can be neglected. Doing so
reduces Eq.~(\ref{eq_H_q}) to the Hamiltonian of a single qubit. The
$N-1$ qubits far from their degeneracy points can be considered as
conventional classical magnetic moments and their influence on the
Nth qubit provides an additional bias $\epsilon_{N,\mathrm{eff}}$
only. The resulting Hamiltonian is $H_q = -
(\epsilon_{N,\mathrm{eff}}+ \epsilon_{N})\sigma_{z}^{(N)} + \Delta_N
\sigma_{x}^{(N)} + \mathrm{Const.}$ Therefore, from measurements
described above, and by making use of Eqs.
~(\ref{dip1}),~(\ref{dip2}), the persistent current $I_{p,N}$ as
well as $\Delta_N$ can be determined. By repeating this procedure
for the other qubits we can find their parameters.

In order to complete the determination of the parameters of $H_q$
the coupling energies $J_{ij}$ must be obtained. They can be
obtained from the shift of the qubit dips relative to the common
degeneracy point. This can be seen in the numerically calculated
energy levels and qubit response in Figs.~\ref{EnergyAFM}
and~\ref{AFresponse}. For low $T$ and small $\Delta_i$, the
locations of the dips in $\tan\Theta(f_1,f_2,\ldots)$, due to
anti-crossings, simply follow the separation lines between the
different classical states in the classical stability diagram. Due
to this the state of the two-qubit system can be easily
reconstructed. Moreover, the coupling energy follows directly from
this stability diagram. For instance, the transition
$\mathopen|\uparrow\uparrow\rangle \longleftrightarrow
\mathopen|\downarrow\uparrow\rangle$ occurs at $\epsilon_1=J$.
Therefore, one can easily show that the coupling energy follows from
$J=\delta f_{q1} \Phi_0 I_{p,1}=\delta f_{q2} \Phi_0 I_{p,2}$.

\begin{figure}\centering
  % Requires \usepackage{graphicx}
  \includegraphics[width=6cm]{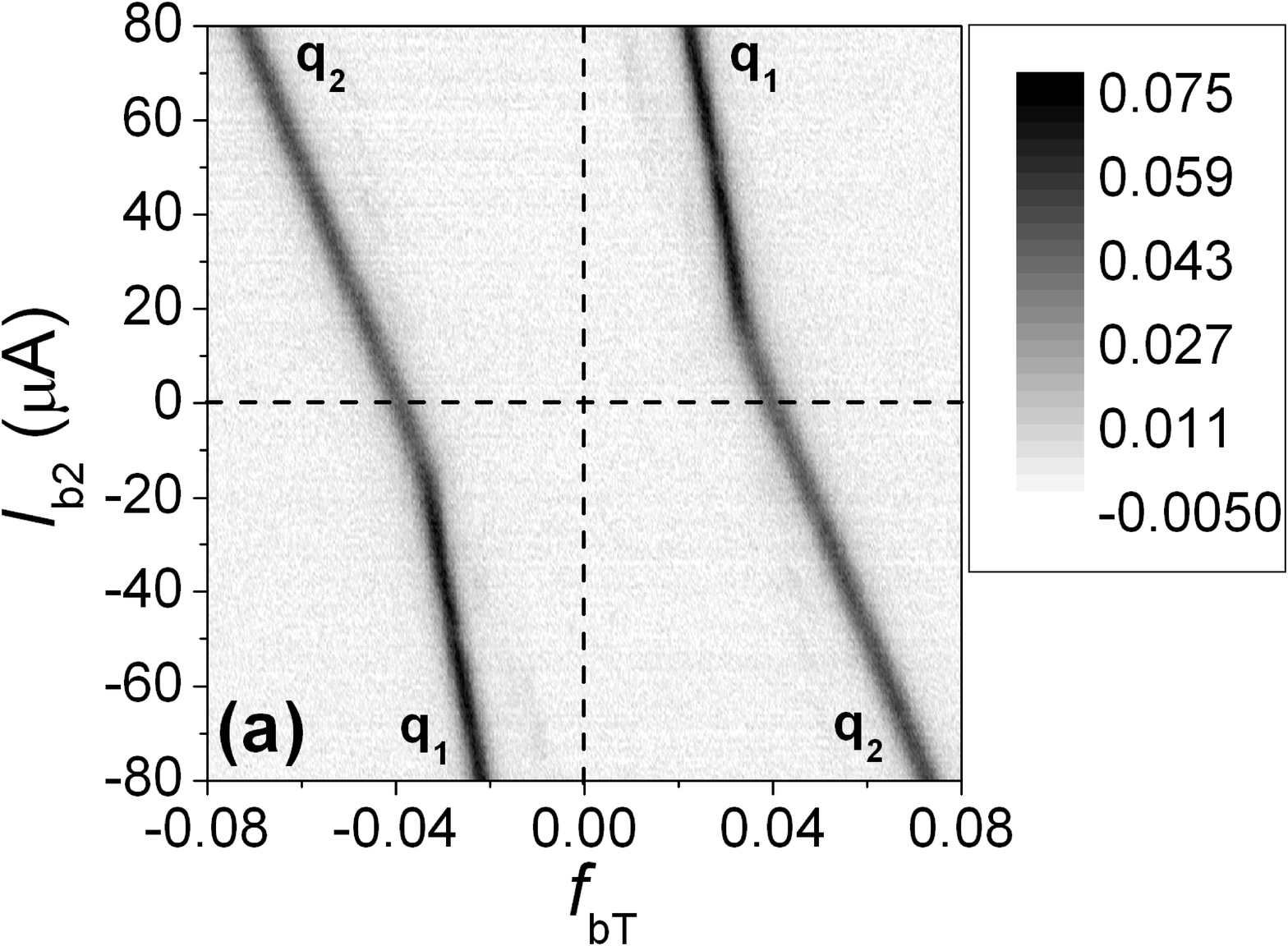}
  \vspace{2mm}
  \includegraphics[width=6cm]{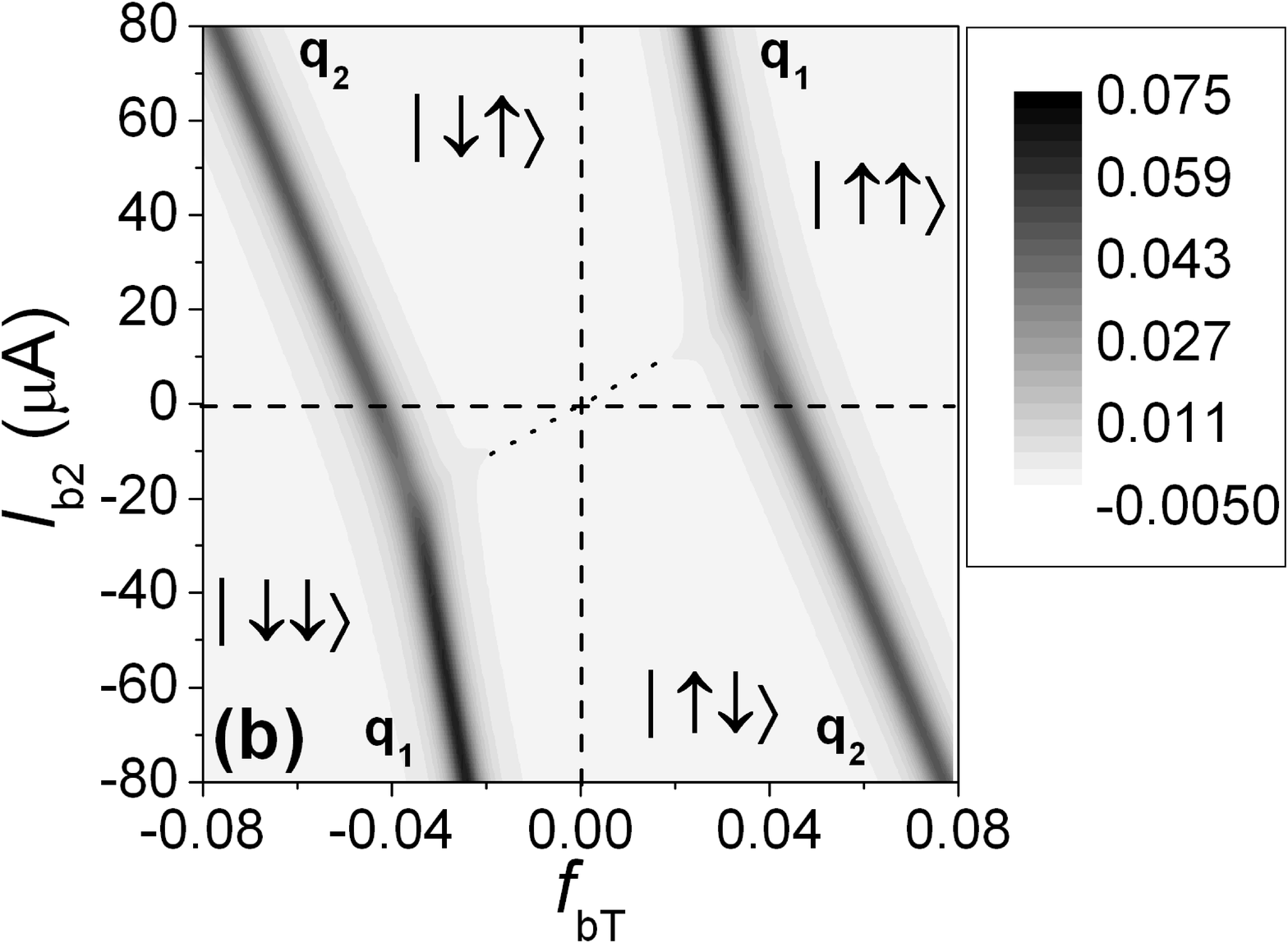}\\
\caption{(a) $-\tan\Theta(I_{b2},I_{bT})$ for AFM-coupled qubits
measured at the mixing chamber temperature of 10 mK. (b) Theoretical
fit for an effective temperature $T_\mathrm{eff}=30$~mK and for the
sample parameters $\Delta_1=45,~\Delta_2=55,~J=550$ (all in mK) and
$I_{p1}=92$~nA, $I_{p2}=84$~nA.}\label{2Q30mK}
\end{figure}

In Fig.~\ref{2Q30mK} we show some quite recent results from a 2
qubit system measured with an improved setup. The qubits in this
sample are coupled by a common Josephson junction as in
Ref.~\cite{Grajcar2005a}. This allows one to design the coupling
strength over a wider range and than with inductive coupling only.
By using the procedure described above all parameters of the
Hamiltonian, Eq.~\ref{eq_H_q}, were found from the experimental
data. The mutual inductances $M_{i T}$ between the coil and qubits
were determined from the periodicity of the qubit signal from a scan
over multiple $\Phi_0$. The mutual inductances between the bias
lines and qubits followed from the slope of the qubit lines in
Fig.~\ref{2Q30mK}a. These mutual inductances can be used to
calculate the energy biases $\epsilon_i$ and, therefore, for the
prediction of the response of the tank from the theory of Sec.
\ref{Sec:theory}. This prediction is shown shown in
Fig.~\ref{2Q30mK}b. The good agreement between experimental data and
the theoretical prediction confirms the systems effective
temperature of about 30~mK which also followed from the fit.

%\begin{figure}
  % Requires \usepackage{graphicx}
%  \includegraphics[width=7cm]{tan_ferro}\\
%\caption{$-\tan\Theta$ for an FM-coupled two qubit system. The
%tank response is calculated for the same values as in
%Fig.~\ref{EnergyFM}, i.e., $I_{p,1}=150~\mathrm{\mu A}$,
%$I_{p,2}=100~\mathrm{\mu A}$, $\Delta_1=\Delta_2=100~\mathrm{mK}$
%and J=-300~mK.}\label{FMresponse}
%\end{figure}

%\begin{figure}
  % Requires \usepackage{graphicx}
%\centering \includegraphics[width=3.5cm]{4qlayout}
 %\includegraphics[width=7cm]{4qexp}\\
%\caption{(a) Layout of four coupled qubits. (b)
%$-\tan\Theta(I_{b1},I_{bT})$ measured at the mixing chamber
%temperature of 10 mK. Dashed circle shows the region where pair of
%qubits interact ferromagnetically, while dotted circles correspond
%to anti-ferromagnetically interacting pairs of qubits.}\label{4Q}
%\end{figure}

\section{Adiabatic Quantum Computation}\label{sec:AQC}

\begin{figure}\centering
  % Requires \usepackage{graphicx}
  \includegraphics[width=6cm]{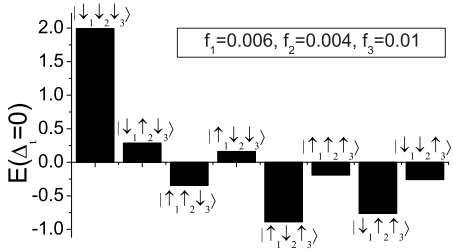}\\
\caption{Energy  of the system for various vectors.
$J_{12}=J_{23}=J_{13}=0.3$~K. $\varepsilon_1=0.315$~K,
$\varepsilon_2=0.252$~K, and $\varepsilon_3=0.525$~K.
$\mathopen|\uparrow_1\downarrow_2\uparrow_3\rangle$ is a global
minimum, while $\mathopen|\downarrow_1\uparrow_2\uparrow_3\rangle$
is a local one.}\label{AQChyst}
\end{figure}

\begin{figure}\centering
  % Requires \usepackage{graphicx}
  \includegraphics[width=6cm]{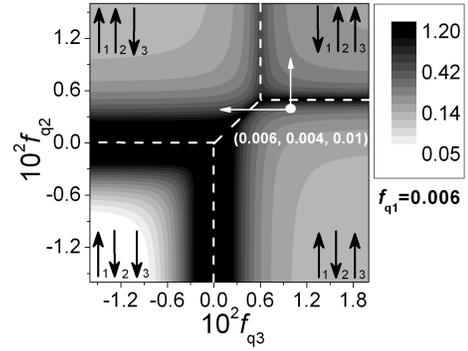}\\
\caption{$-\tan\Theta(f_{q2},f_{q3})$ at $f_{q1}=0.006$. The tank's
response is calculated for $\Delta_1=\Delta_2=\Delta_3=96$~mK,
$J_{12}=J_{13}=J_{23}=300$~mK, $I_{p1}=I_{p3}=350$~nA,
$I_{p2}=420$~nA, and $T=10$~mK. The white dashed lines denote the
cross-overs between the different classical states. At the white dot
$(0.006, 0.004, 0.01)$ the MAXCUT problem with the solution
$\mathopen|\uparrow_1\downarrow_2\uparrow_3\rangle$ is encoded, see
Fig.~\ref{AQChyst}. The solid arrows show the directions in which
the read-out should be carried out in order to reconstruct this
state.}\label{AQCreadoutth}
\end{figure}
In this section we review the AQC-algorithm for the solution of the
MAXCUT problem. In particular we discuss a possible demonstration of
AQC by making use of a system composed of three coupled flux-qubits
using a resonant tank circuit for the readout.

Since the Hamiltonian of Eq.~(\ref{eq_H_q}) is similar to an Ising
Hamiltonian at $\Delta_i=0$ it encodes the  MAXCUT of a 3 vertex
graph if $\Delta_i/J_i \ll 1$ \cite{Grajcar2005}. The simplest
non-trivial case is a three-qubit system as shown in
Fig.~\ref{setup}, for example with the following qubit parameters:
$\Delta_1=\Delta_2=\Delta_3=96$~mK, $J_{12}=J_{13}=J_{23}=300$~mK,
$I_{p1}=I_{p3}=350$~nA and $I_{p2}=420$~nA. If this system is
allowed to adiabatically evolve to the qubit energy biases
$\varepsilon_1(0.006)=0.315$~K, $\varepsilon_2(0.004)=0.252$~K, and
$\varepsilon_3=0.525(0.01)$~K, it will encode a MAXCUT problem with
a solution given by the state
$\mathopen|\uparrow_1\downarrow_2\uparrow_3\rangle$. This can be
seen in Fig.~\ref{AQChyst} where the energies of all flux-states for
the Hamiltonian of Eq.(\ref{eq_H_q}) with all $\Delta_i=0$, $J_{ij}$
and $\varepsilon_{i}$ as given above have been depicted. This system
also exhibits a local minimum
$\mathopen|\downarrow_1\uparrow_2\uparrow_3\rangle$, two spin-flips
away from the global minimum, this makes it an interesting test
system for AQC. If a violation of the adiabatic evolution criterium
occurs, the system can be found in this local minimum instead of the
global one. This can be due to either thermal excitations during the
readout or Landau-Zener transitions caused by a too high readout
speed. Therefore one should optimize this readout speed depending on
the energy gap in order to obtain optimal results.

The tank circuit response $-\tan\Theta(f_{q2},f_{q3})$ predicted by
Eqs. (\ref{eq1_shift}) and (\ref{eq2_shift}) for a fixed value
$f_{q1}=0.006$ is shown in Fig.~\ref{AQCreadoutth}. The white dashed
lines denote the cross-overs between the different classical
($\Delta_i=0$) states, which are marked in the figure by the state
vectors $\mathopen|\downarrow_1\uparrow_2\uparrow_3\rangle$,
$\mathopen|\uparrow_1\uparrow_2\downarrow_3\rangle$,
$\mathopen|\uparrow_1\downarrow_2\downarrow_3\rangle$ and
$\mathopen|\uparrow_1\downarrow_2\uparrow_3\rangle$. The main
feature of the tank's response is that it restores the classical
cross-overs. This fact can be used for reading out the state of the
qubits at any flux point. In the case of Fig.~\ref{AQCreadoutth} we
want to know the qubits' configuration at $(0.006,0.004,0.01)$,
because this point encodes the solution of the MAXCUT problem that
we are interested in. Therefore, if we perform the measurement of
the qubits' susceptibility in the directions marked by solid arrows
in Fig.~\ref{AQCreadoutth}, a \emph{peak} structure will appear
either to the left, or the right side of $(0.006,0.004,0.01)$ in
each flux direction. This allows the establishment of the following
criterium: if the peak appears to the right(left) side of the
starting point when we scan the flux $f_{qi}$ through the qubit $i$,
the starting point corresponds to the classical state $\downarrow_i
(\uparrow_i)$. Therefore, one can easily check that $(0.006, 0.004,
0.01)$ corresponds to the
$\mathopen|\uparrow_1\downarrow_2\uparrow_3\rangle$ state if all
$\Delta_i=0$ (for finding $f_1$ one would need another figure
\textit{e.g.} the $(f_{q1},f_{q3})$ plane at a fixed value of
$f_{q2}=0.004$).

\begin{figure}\centering
\includegraphics[width=5cm]{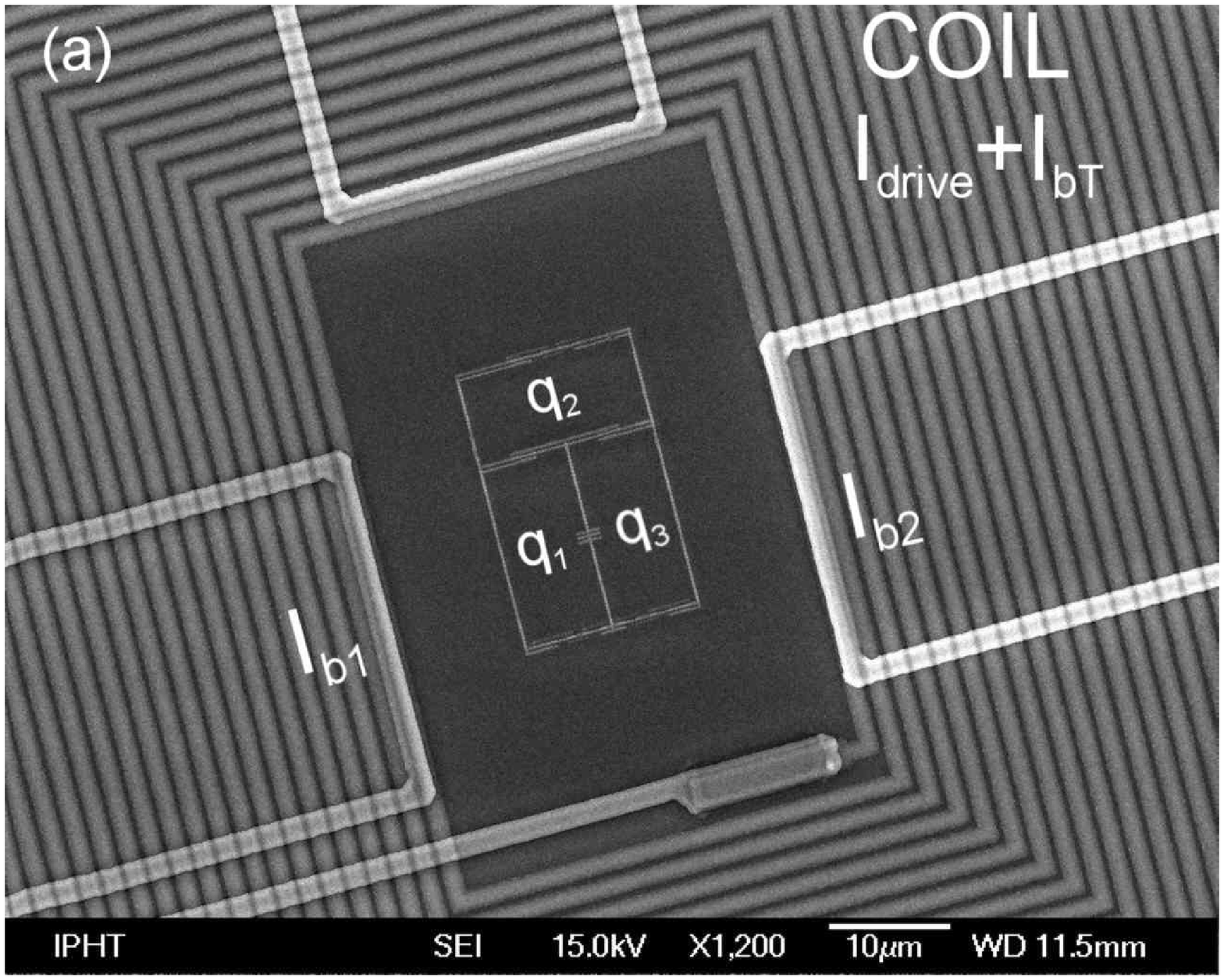}\\
\vspace{2mm}
 \includegraphics[width=6cm]{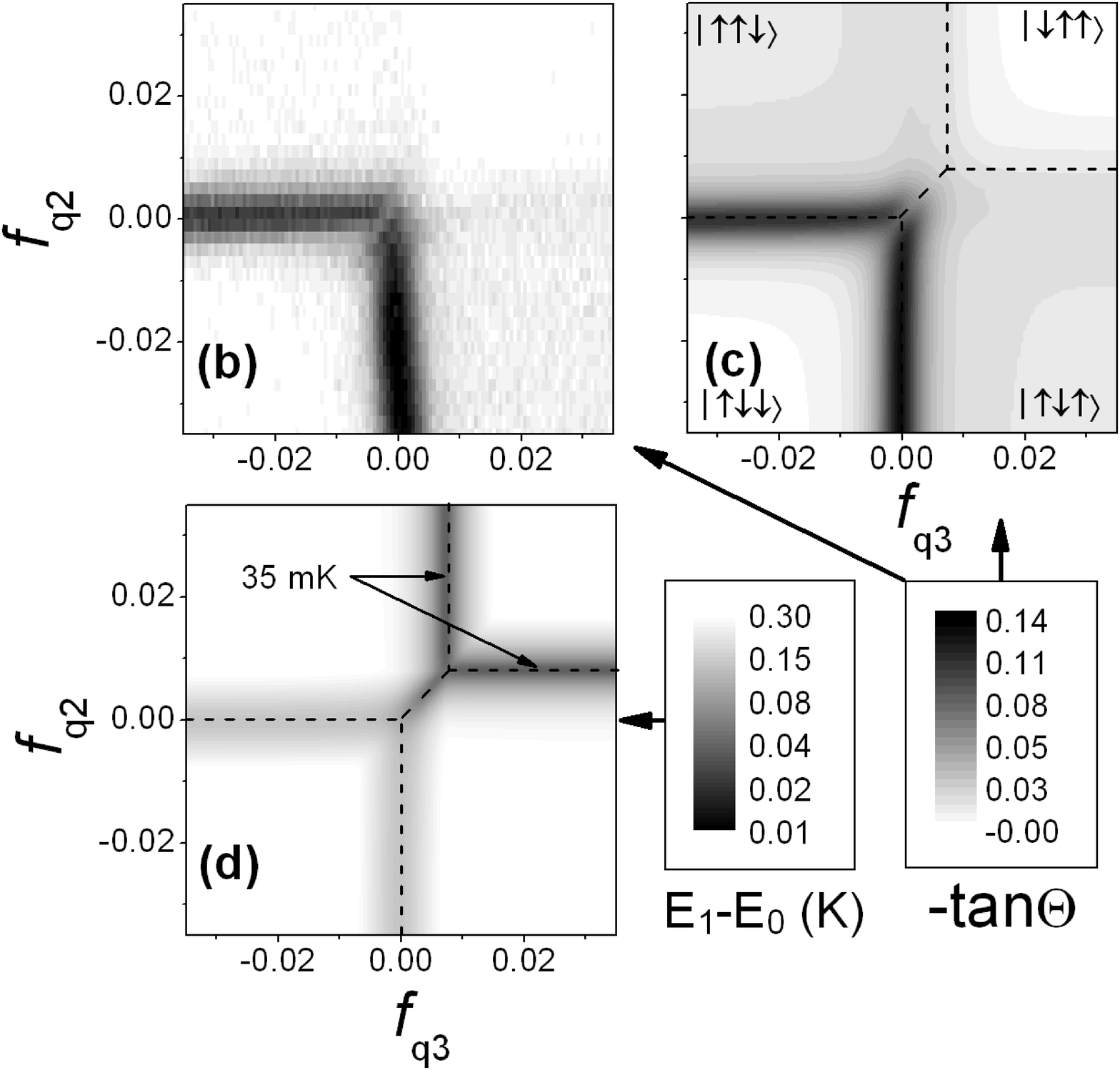}\\
\caption{(a) micrograph of the sample (b)
$-\tan\Theta(f_{q2},f_{q3})$ at $f_{q1}=0.008$, measured at
$T_\mathrm{mix}=10$~mK. (c) predicted response for
$\Delta_1=\Delta_2=\Delta_3=70$~mK, $J_{12}=J_{13}=J_{23}=610$~mK,
$I_{p1}=I_{p2}=115$~nA, $I_{p3}=125$~nA, and $T_\mathrm{eff}=70$~mK.
(d) Difference between the ground and first excited states. For (c)
and (d) the black dashed lines denote the cross-overs between the
different classical states.}\label{AQCreadout}
\end{figure}

As a first step towards the demonstration of AQC we measured the
full three-dimensional susceptibility of three
anti-ferromagnetically coupled qubits \cite{Izmalkov2005}. For this
purpose we fabricated a sample with the layout of Fig.~\ref{setup}
and measured it in our setup at a base temperature of 10 mK.
Figure~\ref{AQCreadout}a shows a micrograph of the sample: three Al
persistent current qubits are fabricated inside a Nb pancake coil.
Two junctions in each qubit are nominally $600\times120$~nm$^2$,
while a third one is $\sim$35\% smaller. Each qubit is coupled to
the other two both magnetically and via shared $120 \times
2000$~nm$^2$ junctions. The flux through the qubits is controlled by
direct currents through the coil $I_{\mathrm{bT}}$, and two
additional lines $I_{\mathrm{b1}}$ and $I_{\mathrm{b2}}$ (the third
line was not used during the experiment). The Nb coil has an
inductance $L_\mathrm{T}=134$~nH, and together with an external
capacitance $C_\mathrm{T}=470$~pF forms a parallel tank circuit with
$\omega_\mathrm{T}/2\pi=20.038$~MHz and quality $Q_\mathrm{T}=700$.
The qubit-coil mutual inductances were extracted from the $\Phi_0$
periodicity of the $ac$-susceptibility of the individual qubits as
$M_{q1\mathrm{T}}\approx 45.8$~pH and $M_{q2\mathrm{T}}\approx
46.6$~pH, $M_{q3\mathrm{T}} \approx 45.8$~pH. The mutual inductances
between all qubits and biasing wires were found from two scans
$(I_{bT},I_{b1}), (I_{bT},I_{b2})$, see Ref.~\cite{Izmalkov2005}.

Fig.~\ref{AQCreadout}b shows $-\tan\Theta(f_{q2},f_{q3})$ at
$f_{q1}=0.008$, while Fig.~\ref{AQCreadout}c is the theoretical
prediction for $\Delta_1=\Delta_2=\Delta_3=70$~mK,
$J_{12}=J_{13}=J_{23}=610$~mK, $I_{p1}=I_{p2}=115$~nA,
$I_{p3}=125$~nA, and $T=70$~mK. The experimental and theoretical
data are found to be in good agreement in the full three-dimensional
flux space (other data for $f_{q1}=0.005,~0,~-0.005,~-0.008$ can be
found in \cite{Izmalkov2005}). However, for this sample not all
classical cross-overs could be reconstructed. The narrow lines in
the upper right quadrant of Fig.~\ref{AQCreadout}bc are absent.
There are two main reasons for this: firstly the difference between
the ground and first excited state ($35$~mK) along these measured
lines is small (see ~Fig.\ref{AQCreadout}d) in comparison with the
effective temperature $T=70$~mK for this measurements. Secondly all
qubits' persistent currents (or magnetic moments) are similar,
therefore there is no significant \emph{magnetization} change along
the transitions $\mathopen|\uparrow_1\uparrow_2\downarrow_3\rangle
\longleftrightarrow
\mathopen|\downarrow_1\uparrow_2\uparrow_3\rangle$ and
$\mathopen|\downarrow_1\uparrow_2\uparrow_3\rangle$
$\longleftrightarrow
\mathopen|\uparrow_1\downarrow_2\uparrow_3\rangle$, thus there is no
susceptibility change and no phase shift. In order to observe these
lines, one could make a sample with different persistent currents,
increase the gap between the ground and first excited states, or
decrease the system's effective temperature. For the latter approach
we already made some progress because for our improved experimental
setup, see~Fig.\ref{2Q30mK}, we have reached an effective
temperature $\sim30$~mK.

The next steps for AQC will be a improving the readout speed, the
demonstration of the real adiabatic evolution/computation and
demonstration of its efficiency in comparison with the classical
simulated annealing.

\section{Discussion}

This review is concluded by answering some general questions that we
have been often asked. The first question is whether it is possible
to realize AQC with coupled classical magnetic moments? Or in other
words: do we need quantum mechanics for an AQC implementation? In
the quantum system there is an energy gap between the ground and the
first exited states, thus the qubit system can evolve always staying
in the ground state. A system without level- anticrossings exhibits
many metastable states, so that evolution in the ground state is not
possible.

As the qubits stability diagrams, shown in
Figs.~\ref{AFresponse},~\ref{AQCreadoutth} are the same as would be
is expected for coupled classical magnetic moments, where is the
proof that the system described above behaves quantum mechanically?
In the classical limit the persistent current ``qubits'' have a
magnetic hysteresis for any finite qubit's inductance. This
hysteresis was observed by making use of our method~\cite{iliAPL}.
In this work instead of hysteresis we observed a narrow dip, which
reflects the level anticrossing due to quantum tunneling. Therefore,
there is a nonzero gap between the ground and the first exited state
which allows AQC.

The stability of AQC against external noise, to our knowledge, is
not known. On one hand, since we do not use any coherent
oscillations, AQC should not be very sensitive to the environment.
On the other hand, external noise can change the Hamiltonian itself
and, therefore, its ground state. This issue requires further study.

In the present work we discussed our approach for demonstrating an
adiabatic quantum algorithm by making use of a coupled flux-qubit
system. By making use of a low-frequency resonator inductively
coupled to the qubits we can completely reconstruct the parameters
of a multi-qubit system. This multi-qubit ground state anticrossings
read-out can be currently performed at an effective temperature of
30~mK. The experimental data are found to be in complete agreement
with quantum mechanical predictions. We have reconstructed the
susceptibility of three coupled flux qubits in full parameter space.
The next steps towards AQC will be improvement of the read-out speed
and the demonstration of adiabatic quantum computation and its
efficiency.

\appendix[Susceptibility calculation]\label{suceptibility} Starting
from the first order approximation of the total flux given in the
main text (eq. \ref{eq2_Phi_Tot}) we can write out the commutator as
a sum of the qubit response functions $\varphi_{ij}$:
\begin{equation} \label{eq1_Resp_Qub}
\langle (i/\hbar)[\Phi(t),\Phi(t_1)]_-\rangle \theta(t-t_1) =
\sum_{ij} \lambda_i \lambda_j \varphi_{ij}(t,t_1).
\end{equation}
with
\begin{equation}
\label{eq2_Resp_Qub} \varphi_{ij}(t,t_1) = \langle
(i/\hbar)[\sigma_{z}^{(i)}(t),\sigma_{z}^{(j)}(t_1)]_-\rangle
\theta(t-t_1),
\end{equation}
where we introduced the Heaviside step function: $\theta(t-t_1) =
1,$ if $ t>t_1,$ and  $\theta(t-t_1) = 0,$ when $ t<t_1.$

For small values of the tank current and voltage the derivative of
the total qubit flux is given by the expression
\begin{equation} \label{eq_der_Flux}
\frac{d}{dt} \langle \Phi_{tot}(t) \rangle =\sum_{ij}
\frac{\lambda_i \lambda_j}{L_T} \int dt_1 \varphi_0(t,t_1) \langle
V_T \rangle (t_1).
\end{equation}
Putting this into Eq. (\ref{eq1_V_T}) results in a shift of the
resonant frequency of the tank which is proportional to the
susceptibility and is just the Fourier transform of the response
function:
\begin{eqnarray} \label{eq1_Chi}
\chi_{ij}(\omega) = \int d\tau e^{i\omega \tau} \varphi_{ij}(\tau).
\end{eqnarray}
This shift in the resonance frequency causes the phase shift when
the tank is driven with its unloaded resonance frequency $\omega_T$
according to Eq. (\ref{eq1_shift}).

In order to calculate the linear response functions
$\varphi_{ij}(t,t_1)$ (Eq. (\ref{eq2_Resp_Qub})) we have to find the
product of two projections of Pauli matrices taken at times $t$ and
$t_1$
\begin{equation} \label{eq1_prod}
\langle \sigma_{z}^{(i)}(t)\sigma_{z}^{(j)}(t_1)\rangle =
\sum_{\mu}\rho_{\mu} \langle \mu
|\sigma_{z}^{(i)}(t)\sigma_{z}^{(j)}(t_1)|\mu \rangle,
\end{equation}
averaged over the quantum-mechanical states $\langle
\mu|..|\mu\rangle$ and the equilibrium distribution $\rho_{\mu} =
e^{-E_{\mu}/k_B T}/(\sum_{\alpha}e^{-E_{\alpha}/k_B T})$.

As a result we obtain the expression
\begin{eqnarray} \label{eq1_resp}
\varphi_{ij}(t,t_1) = \varphi_{ij}(t-t_1) = (i/\hbar)
\theta(t-t_1)\nonumber\\
 \sum_{\mu}\rho_{\mu} \sum_{\beta} \{
 \langle \mu | \sigma_{z}^{(i)} |\beta \rangle \langle \beta | \sigma_{z}^{(j)} |\mu \rangle
e^{i\omega_{\mu \beta }(t-t_1)}-\nonumber\\
\langle \mu | \sigma_{z}^{(j)} |\beta \rangle \langle \beta |
\sigma_{z}^{(i)} |\mu \rangle  e^{i\omega_{\beta \mu}(t-t_1)} \},
\end{eqnarray}
for the linear response functions $\varphi_{ij}(t,t_1)$
(\ref{eq1_resp}). which give after Fourier transformation:
\begin{eqnarray} \label{eq2_Chi}
\chi_{ij}(\omega) = - \sum_{\mu \nu} \frac{\rho_{\mu} - \rho_{\nu}}
{\hbar \omega + E_{\mu} - E_{\nu} +i0}\nonumber\\ \langle \mu
|\sigma_{z}^{(i)} |\nu \rangle \langle \nu | \sigma_{z}^{(j)} |\mu
\rangle .
\end{eqnarray}
By taking into account that the resonant frequency of the tank is
much smaller than the level spacing and, therefore, can be neglected
Eq. (\ref{eq2_shift}) of the main text is generated.

% use section* for acknowledgement
\section*{Acknowledgment}
We thank D.~V.~ Averin, A.~Blais,  A.~Shnirman, E.~Goldobin,
Ya.~S.~Greenberg, R.~Gross, H.~E.~Hoenig, Yu.~A.~Pashkin,
M.~J.~Storcz, F.~K.~Wilhelm and A.~Zeilinger for fruitful
discussions.

% references section
% NOTE: BibTeX documentation can be easily obtained at:
% http://www.ctan.org/tex-archive/biblio/bibtex/contrib/doc/

% can use a bibliography generated by BibTeX as a .bbl file
% standard IEEE bibliography style from:
% http://www.ctan.org/tex-archive/macros/latex/contrib/supported/IEEEtran/bibtex
\bibliographystyle{IEEEtran}
% argument is your BibTeX string definitions and bibliography database(s)

%\bibliography{references}

\begin{thebibliography}{10}
\providecommand{\url}[1]{#1} \csname url@rmstyle\endcsname
\providecommand{\newblock}{\relax} \providecommand{\bibinfo}[2]{#2}
\providecommand\BIBentrySTDinterwordspacing{\spaceskip=0pt\relax}
\providecommand\BIBentryALTinterwordstretchfactor{4}
\providecommand\BIBentryALTinterwordspacing{\spaceskip=\fontdimen2\font
plus \BIBentryALTinterwordstretchfactor\fontdimen3\font minus
  \fontdimen4\font\relax}
\providecommand\BIBforeignlanguage[2]{{%
\expandafter\ifx\csname l@#1\endcsname\relax
\typeout{** WARNING: IEEEtran.bst: No hyphenation pattern has been}%
\typeout{** loaded for the language `#1'. Using the pattern for}%
\typeout{** the default language instead.}%
\else \language=\csname l@#1\endcsname \fi #2}}

\bibitem{Farhi2000}
E.~Farhi, J.~Goldstone, S.~Gutmann, and M.~Sipser, ``Quantum
computation by
  adiabatic evolution,'' quant-ph/0001106.

\bibitem{Bertet2005b}
P.~Bertet, I.~Chiorescu, G.~Burkard, K.~Semba, C.~J. P.~M. Harmans,
D.~P.~DiVincenzo, and J.~E.~Mooij, ``Dephasing of a superconducting
qubit induced by photon noise,'' \emph{Phys. Rev. Lett}, vol.~95,
no.~25, pp. 257002, 2005.

\bibitem{Yoshihara2006}
F.~Yoshihara, K.~Harrabi, A.~O. Niskanen, Y.~Nakamura, and J.~S.
Tsai,
  ``Decoherence of flux qubits due to 1/f flux noise,'' \emph{Phys. Rev. Lett}, vol.~97, no.~16, pp. 167001, 2006.

\bibitem{Kaminsky2004}
W.~M. Kaminsky, S.~Lloyd, and T.~P. Orlando, ``Scalable
superconducting
  architecture for adiabatic quantum computation,'' quant-ph/0403090.

\bibitem{Mooij1999}
J.~E. Mooij, T.~P. Orlando, L.~Levitov, L.~Tian, C.~H. van~der Wal,
and
  S.~Lloyd, ``Josephson persistent-current qubit,'' \emph{Science}, vol. 285,
  pp. 1036--1039, 1999.

\bibitem{Steffen2003}
M.~Steffen, W.~van Dam, T.~Hogg, G.~Breyta, and I.~Chuang,
``Experimental
  implementation of an adiabatic quantum optimization algorithm,'' \emph{Phys.
  Rev. Lett}, vol.~90, no.~6, pp. 067903, 2003.

\bibitem{Garey1976}
M.~R. Garey, D.~S. Johnson, and L.~Stockmeyer,``Some simplified
NP-complete graph problems,''  \emph{Theor. Comput. Sci.},
  vol.~1, pp. 237--267, 1976.

\bibitem{Grajcar2005}
M.~Grajcar, A.~Izmalkov, and E.~Il'ichev, ``Possible implementation
of
  adiabatic quantum algorithm with superconducting flux qubits,'' \emph{Phys.
  Rev. B}, vol.~71, pp. 144501, 2005.

\bibitem{Izmalkov2004a}
A.~Izmalkov, M.~Grajcar, E.~Il'ichev, T.~Wagner, H.-G. Meyer, A.~Y.
Smirnov,
  M.~H.~S. Amin, A.~M. van~den Brink, and A.~M. Zagoskin, ``Evidence for
  entangled states of two coupled flux qubits,'' \emph{Phys. Rev. Lett},
  vol.~93, pp. 037003, 2004.

\bibitem{Majer2005}
J.~B. Majer, F.~G. Paauw, A.~C.~J. ter Haar, C.~J. P.~M. Harmans,
and J.~E.
  Mooij, ``Spectroscopy on two coupled superconducting flux qubits,''
  \emph{Phys. Rev. Lett}, vol.~94, no.~9, pp. 090501, 2005.

\bibitem{Levitov2001}
L.~S. Levitov, T.~P. Orlando, J.~B. Majer, and J.~E. Mooij,
``Quantum spin
  chains and {Majorana} states in arrays of coupled qubits,'' cond-mat/0108266.

\bibitem{Grajcar2005a}
M.~Grajcar, A.~Izmalkov, S.~H.~W. van~der Ploeg, S.~Linzen,
E.~Il'ichev, T.~Wagner, U.~Hubner, H.-G. Meyer, A.~Maassen van~den
Brink, S.~Uchaikin, and A.~M. Zagoskin, ``Experimental realization
of direct josephson coupling between superconducting flux qubits,''
\emph{Phys. Rev. B}, vol.~72, no.~2, pp. 020503, 2005.

\bibitem{Ploeg2006}
S.~H.~W. van~der Ploeg, A.~Izmalkov, A.~Maassen van~den Brink,
U.~H\"ubner,
  M.~Grajcar, E.~Il'ichev, H.-G. Meyer, and A.~M. Zagoskin, ``Controllable
  coupling of superconducting flux qubits,'' \emph{Phys. Rev. Lett}, vol.~98, no.~5, pp. 057004, 2007.

\bibitem{Ilichev2004}
E.~Il'ichev, A.~Y. Smirnov, M.~Grajcar, A.~Izmalkov, D.~Born,
N.~Oukhanski,
  T.~Wagner, W.~Krech, H.-G. Meyer, and A.~Zagoskin, ``Radio-frequency method
  for investigation of quantum properties of superconducting structures,''
  \emph{Fiz. Nizk. Temp.}, vol.~30, no.~30, pp. 823--833, 2004.

\bibitem{May2003}
T.~May, E.~Il'ichev, H.-G. Meyer, and M.~Grajcar, ``Microfabricated
oscillator for radio-frequency microscopy with integrated magnetic
field concentrator,''
  \emph{Rev. Sci. Instr.}, vol.~74, no.~3, pp. 1282--1284, 2003.

\bibitem{Greenberg2002}
Y.~S. Greenberg, A.~Izmalkov, M.~Grajcar, E.~Il'ichev, W.~Krech,
H.-G. Meyer, M.~H.~S. Amin, and A.~Maassen van~den Brink,
``Low-frequency characterization of quantum tunneling in flux
qubits,'' \emph{Phys. Rev. B}, vol.~66, no.~21, pp. 214525, 2002.

\bibitem{Izmalkov2005}
A.~Izmalkov, M.~Grajcar, S.~H.~W. van~der Ploeg, U.~H\"ubner,
E.~Il'ichev, H.-G. Meyer, and A.~M. Zagoskin, ``Measurement of the
ground-state flux diagram of three coupled qubits as a first step
towards the demonstration of adiabatic quantum computation,''
\emph{Europhys. Lett.}, vol.~76, no.~3, pp. 533--539, 2006.

\bibitem{iliAPL}
E.~{Il'ichev}, Th.~Wagner, L.~Fritzsch, J.~Kunert, V.~Schultze,
T.~May, H.~E.~Hoenig, H.~G.~Meyer, M.~Grajcar, D.~Born, W.~Krech
M.~V.~Fistul A.~M.~Zagoskin, ``Characterization of superconducting
structures designed for qubit realizations,'' \emph{Appl. Phys.
Lett}, vol. 80, pp. 4184-4186, 2002.

\end{thebibliography}

\end{document}